\documentclass[runningheads]{llncs}
\usepackage[T1]{fontenc}
\usepackage{amsmath}  
\usepackage{amssymb}  
\usepackage{enumitem}
\usepackage{xcolor}
\usepackage{graphicx}
\usepackage{url}
\usepackage{hyperref}

\usepackage{graphicx}
\usepackage{multicol,multirow}
\newcommand{\orcid}[1]{\href{https://orcid.org/#1}{\includegraphics[width=8pt]{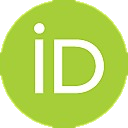}}}
\begin{document}
\title{exHarmony: Authorship and Citations for Benchmarking the Reviewer Assignment Problem}
\titlerunning{exHarmony: Benchmarking the Reviewer Assignment Problem}
%
\author{Sajad Ebrahimi$^*$ \inst{1} \orcid{0009-0003-1630-3938} \and
Sara Salamat$^*$ \inst{1} \orcid{0009-0007-3676-6023}\and
Negar Arabzadeh\inst{1} \orcid{0000-0002-4411-7089} \and
Mahdi Bashari\inst{1} \orcid{0000-0002-9211-5475}\and
Ebrahim Bagheri\inst{1,2} \orcid{0000-0002-5148-6237}}

\authorrunning{S. Ebrahimi, S. Salamat et al.}
\institute{Reviewerly, Toronto ON, Canada
\url{https://reviewer.ly}
\email{\{sajad,sara,negara,bashari,bagheri\}@reviewer.ly}\and
University of Toronto, Toronto ON, Canada
}

\maketitle              
\def\thefootnote{*}\footnotetext{These authors contributed equally to this work.}
\begin{abstract}
The peer review process is crucial for ensuring the quality and reliability of scholarly work, yet assigning suitable reviewers remains a significant challenge. Traditional manual methods are labor-intensive and often ineffective, leading to nonconstructive or biased reviews. This paper introduces the \texttt{exHarmony} (\textit{eHarmony} but for connecting \textbf{ex}perts to manuscripts) benchmark, designed to address these challenges by re-imagining the Reviewer Assignment Problem (RAP) as a retrieval task. Utilizing the extensive data from OpenAlex, we propose a novel approach that considers a host of signals from the authors, most similar experts, and the citation relations as potential indicators for a suitable reviewer for a manuscript. This approach allows us to develop a standard benchmark dataset for evaluating the reviewer assignment problem without needing explicit labels. We benchmark various methods, including traditional lexical matching, static neural embeddings, and contextualized neural embeddings, and introduce evaluation metrics that assess both relevance and diversity in the context of RAP. Our results indicate that while traditional methods perform reasonably well, contextualized embeddings trained on scholarly literature show the best performance. The findings underscore the importance of further research to enhance the diversity and effectiveness of reviewer assignments.

\end{abstract}
%
%
%

\section{Introduction}
The peer review process is a fundamental aspect of academic publishing, ensuring the quality and reliability of scholarly work \cite{arabzadeh2024reviewerly}. Nonetheless, finding suitable reviewers for submitted manuscripts has always been a substantial challenge for publishers, editors, and conference organizers \cite{mittal2019understanding}. Conventional manual methods, such as requesting reviewers to bid on papers, are labor-intensive and often less effective. This method places a significant burden on reviewers, and in cases of unbalanced bidding between reviewers and papers, it can lead to nonconstructive or biased reviews if the reviewers' expertise does not align closely with the paper's content. Furthermore, in contexts like journal review processes, where submission deadlines are distributed throughout the year, the bidding approach would be impractical. Additionally, the time required to assign reviewers can considerably lengthen the submission-to-publication timeline, delaying the dissemination of important research outcomes.

Despite the critical nature of the Reviewer Assignment Problem (RAP), only a few efforts \cite{aksoy2023reviewer,hartvigsen1999conference,kolasa2011survey,wang2010comprehensive,jovanovic2023reviewer} have made attempts to automate parts of the process, and comprehensive, widely adopted solutions remain scarce. While some attempts have been made by industry researchers, this gap in the research underscores the need for innovative, open-source approaches to improve the efficiency and effectiveness of reviewer assignments.
One of the main reasons the task has not been well studied is because data for reviewers are usually not available due to reasons like the anonymity of the reviewers and privacy issues.

In this paper, we aim to acknowledge the challenges faced in RAP and propose \texttt{exHarmony} (a name inspired by 'eHarmony', in connecting experts with manuscripts) benchmark to facilitate research on this topic. By leveraging the extensive data from OpenAlex~\cite{priem2022openalex}, an open database of scholarly works, we provide a benchmark for this understudied task. We leverage a diverse set of entities, including papers, topics, institutions, and authors, to gain a multi-dimensional understanding of the task. We redefine the task by using \textit{weakly supervised labels}, considering both the authors of a paper and the authors of papers cited within it, as potential reviewers. Consequently, to have a fair assessment of the RAP task, we re-frame the task as retrieving those authors based on their previous work as potential reviewers. This allows us to evaluate and improve the task without having explicit labels. 

To avoid making strong assumptions on our ground truths, we developed three distinct sets of ground truth reviewers in our proposed dataset. We introduce \texttt{exHarmony-Authors}, \texttt{exHarmony-Cite}, and \texttt{exHarmony-SimCite}, which consider, respectively, the authors of the paper under review, the authors of the papers cited by the paper under review, and the authors of the top-k most similar cited papers as the ground truth reviewers. Additionally, to refine these sets, we apply a \textit{filter} for established authors, including only those with at least $N$ published papers, to ensure that the reviewers are considered experts in their fields. We will compare and discuss the differences between all six versions of these datasets (both filtered and unfiltered) further in the paper.

Furthermore and for the sake of benchmarking, we provide a set of baselines for the task, including traditional lexical matching approaches \cite{peng2017time}, static neural embedding-based approaches \cite{ogunleye2017proposed}, and fine-tuned pre-trained language models for academic recommendations \cite{beltagy2019scibert,cohan2020SPECTER}. However, our findings indicate that none of the current open-source approaches deliver satisfactory results for this task in real-world scenarios, highlighting the need for further exploration and more in-depth research.
We distinguish our work from previous works, as we conceptualize RAP as an information retrieval (IR) task to leverage established IR techniques for efficiently matching papers with the most relevant reviewers. In this context, papers are treated as queries, and reviewers as items to be ranked.

The recommendation of the reviewer goes beyond simply finding relevant reviewers. As such, in our evaluations, we consider diversity in several aspects. We propose a set of novel evaluation metrics to ensure that the recommended reviewers include individuals at different stages of their careers and from various institutional backgrounds, avoiding biases and promoting a more equitable review process. Therefore, evaluation is conducted based on the relevance and diversity of the retrieved set of reviewers.

In summary, our contributions are as follows:
\begin{itemize}
    \item  We redefine the Reviewer Assignment Problem in a way that it can be effectively studied without requiring \textit{explicit} labeled data on reviewers. 
        \item  We curate \texttt{exHarmony}, a large dataset for the RAP task with three subsets of \texttt{exHarmony-Authors}, \texttt{exHarmony-Cite}, and \texttt{exHarmony-SimCite} and make them publicly available at : \url{https://github.com/sadjadeb/exHarmony}
    \item We provide a set of evaluation metrics that consider relevancy of the reviewers'  expertise with the paper as well as diversity of the set of reviewers.
    \item We provide the results of set of baselines for RAP task on our benchmark, highlighting challenges and suggesting directions for future research.
\end{itemize}

\section{Related Work}

\textbf{Problem Formulations.}
RAP has been modeled in various ways, such as classification task \cite{Zhao2018paperreviewerrecommnedation,Zhang2020multilabel}, topic coverage task \cite{kou2015weighted}, and recommendation \cite{DBLP:journals/corr/abs-0906-4044,peng2017time}. For instance, Zhao et al. \cite{Zhao2018paperreviewerrecommnedation} redefined RAP as a classification task, applying the word mover’s distance method for similarity calculations and the constructive covering algorithm to concurrently classify reviewers and manuscripts. Zhang et al. \cite{Zhang2020multilabel} approached the task as a multilabel classification problem, assigning reviewers based on multiple predicted labels. Conversely, some works have introduced the problem as a topic coverage task \cite{kou2015weighted}.
We suggest modeling RAP as a conventional IR task. In a typical IR task, the objective is to find pertinent documents from a vast collection based on a specific query. Here, the queries are the submitted papers for review, characterized by their metadata. 
The task is essentially reformulated as finding authors whose earlier works closely match the content of the submitted papers.

\textbf{Solutions.}
A few efforts have already been undertaken to tackle the reviewer assignment problem. Those methods that have already been deployed in industrial settings are often proprietary in nature and hence lack transparency, making evaluation and comparison of their performance challenging. Other more transparent solutions for handling RAP typically fall into three categories: topic-based modeling techniques, lexical matching methods, and neural network-based embeddings.
Dumais and Nielsen \cite{Susan1992automatedrap} were the first to tackle RAP as an information retrieval challenge, using the latent semantic indexing (LSI) model to link reviewers with papers. With advancements in topic modeling, Mimno and McCallum \cite{Mimno2007ExpertiseMF} adopted the more sophisticated topic modeling algorithms to introduce the author-persona-topic model to more accurately capture the topics a reviewer might cover. Methods for topic modeling, such as Latent Dirichlet Allocation (LDA) \cite{Mimno2007ExpertiseMF,kou2015weighted}, have been extensively employed for reviewer assignment. These techniques aim to discern the hidden topics within a paper and align them with reviewers who possess knowledge in those domains. Kou et al. \cite{kou2015weighted} implemented topic-weighted coverage calculation using LDA features and introduced the branch-and-bound algorithm to identify reviewers rapidly.
Lexical matching strategies, like TF-IDF \cite{peng2017time}, emphasize extracting statistical characteristics from the texts of reviewers and papers to determine the best match. With the rise of neural-based embeddings, efforts have been made to address RAP using both static word2vec and contextualized embeddings \cite{ogunleye2017proposed,devlin2018bert,mikolov2013efficient}. 
Since we are proposing to redefine RAP as an IR problem, in this paper, we focus on reporting IR-based approaches as benchmarks. 

\textbf{Datasets.}
One of the major obstacles in RAP is the shortage of diverse and large datasets for both training and evaluating current solutions. This lack is partly due to the need to preserve the anonymity of reviewers. In addition, available datasets tend to have a limited range, not having high coverage on different subjects. Moreover, maintaining these datasets is crucial, as they need continuous updates to reflect the latest research developments and reviewer profiles. To the best of our knowledge, there is no dataset available that offers a varied and comprehensive ground truth for RAP. 
Furthermore, there were major problems with previous works \cite{karimzadehgan2008multi,karimzadehgan2009constrained,tang2010expmatching,kou2015weighted,kou2015topicbased,Xu2020Strategyproof,mirzaei2019multiaspect,Nguyen2018ADS}; they all used some conference assignments as the ground truth evaluation data and in fact, those assignments had been done by other RAP solutions whose performances are  unknown and potentially questionable. In other words, earlier works have been evaluated against some unverified anchor. Existing studies that align reviewers with accepted papers as gold standards are few and narrow in scope, often biased towards accepted papers \cite{Xu2020Strategyproof}. Therefore, in this work, we propose the \texttt{exHarmony} dataset, which overcomes the aforementioned challenges by not requiring the collection of reviewer data; instead, we only adopt meta-data from papers, e.g., having paper's title, abstract, their list of authors, and their citations. Since the only requirement for \texttt{exHarmony} is a collection of papers, maintaining it is very easy. It only requires updates with new papers, making it flexible to maintain and keep updated based on author profiles and new research topics.
This characteristic of \texttt{exHarmony} allows it to be large-scale at a very low cost for training and evaluation purposes. This data collection strategy allows us to have data on a diverse set of topics since the pipeline is not limited to data annotation and thus can be applied effortlessly on different topics.
    
\section{The Reviewer Assignment Problem}


We define the reviewer assignment problem as finding a set of reviewers for a given paper whose expertise not only matches the paper's topic but also ensures that the set is diverse, allowing for different opinions and assessments from multiple perspectives. 
In this work, we propose redefining the RAP by hypothesizing that the authors of a paper are potentially the best reviewers for that paper  had they not been its authors. Thus, if one can accurately identify an author as a hypothetical reviewer based solely on their previous publications, the core challenge of RAP could be addressed. However, this assumption presents challenges: 1) the gold standard might be too sparse \cite{arabzadeh2022shallow}, 2) not all authors contribute equally to a paper, and 3) some co-authors, especially early-career researchers, may not yet be ready for independent peer review. 

To mitigate these issues, we explore alternative approaches in order to create a more robust gold standard for reviewers by: 1)  expanding the reviewer pool by including authors of papers that have been cited by the work in question; 2)  considering only authors of top-k most similar cited papers to ensure that the selected cited papers share similar topics with the paper under review; and, 3) constructing a set of  expert reviewers consisting of authors who have published more than $N$ papers i.e., \textit{established authors}, in the past.

Based on these strategies, we propose various variations for \texttt{exHarmony} as follows: (i) considering the authors of a paper as its best potential reviewers (\texttt{exHarmony-Authors}); (ii) treating the authors of the papers that have been cited in the paper under consideration as suitable reviewers (\texttt{exHarmony-Cite}); and (iii) selecting the authors of highly similar cited papers as appropriate reviewers (\texttt{exHarmony-SimCite}). From within these three sets, we define two subsets in order to designate authors with a longer history of publications. As such, we will include both `all authors' as well as `established authors' subsets for each of these three sets. 
Formally, let $P$ be a paper for which we are interested to find reviewers at time $\tau$, with title and abstract $P_t$, and a list of $n$ authors $\mathcal{A}(P) = \{P_{a_1}, P_{a_2}, \ldots, P_{a_n}\}$. We create a collection of authors $C_\tau$ from all the authors of papers published before time $\tau$ as:
\begin{equation}
 C_\tau = \{P_{a_i} \forall P_{a_i} \in \mathcal{A}(P) \mid  \mathcal{T}(P) < \tau \} 
\end{equation}
\noindent where the function $\mathcal{T}$ indicates the publication or submission time of $P$ in the collection. The reviewers are represented by their individual previous papers published prior to time $\tau$. As such, one author can be represented as individual items indexed in the collection by different papers, e.g., $P_{a_i}$ and $P'_{a_j}$ could be the same reviewer coauthoring on both papers $P$ and $P'$.
Therefore, in the author retrieval step, the goal is to retrieve the most relevant and similar previous works $\mathcal{F}(P_t, C_\tau)$ using a retriever $\mathcal{F}$ from the collection of authors (reviewers) $C_\tau$ given the paper title and abstract $P_t$. We consider authors of the similar previous work as the potential pool of reviewers $\mathcal{R}_P$. The set of initial authors $\mathcal{R}_P = \mathcal{F}(P_t, C_\tau)$ might contain repeated reviewers represented by different papers since an author could be retrieved from different papers. 
The curated pool of candidate reviewers is likely to be redundant, with some authors appearing multiple times due to their contributions to different relevant papers. In this step, the goal is to find a ranked list of unique, non-repeated reviewers from the initial pool $\mathcal{R}_P$. We believe that the aggregation step alone has significant potential for future research, as it allows us to balance the trade-off between diversity and expertise or relevance of the suggested set of reviewers.
Thus, the goal of $AGG(\mathcal{R}_P)$ will be to return a ranked list of the top $k$ unique reviewers $[r_1, r_2, \ldots, r_k]$ from the pool $\mathcal{R}_P$. 

\section{The \texttt{exHarmony} Dataset}

In this section, we will elaborate on our proposed dataset, referred to as \texttt{exHarmony}. First, we describe how we collected the data, followed by an overview of the statistics and details of the dataset and its ground truths.

\subsection{Data Collection and Statistics}
We leverage OpenAlex\footnote{\url{https://openalex.org/}} \cite{priem2022openalex} as the foundation for our dataset. OpenAlex is a comprehensive open database of scholarly works and  provides metadata on a wide range of academic articles, including their authors, affiliations, topics, and citation relationships. This extensive coverage allows \texttt{exHarmony} cover many different fields.
The data contains detailed information about papers, such as but not limited to, authors, and affiliations. We have implemented a data pipeline that imports the raw data to MongoDB which provides a platform to query the data, create slices of the dataset, and store the relevant information in \texttt{exHarmony}.
Specifically, we used the March 2024 snapshot of OpenAlex. To make it feasible for researchers in academia with limited computational resources to work with the data, we narrowed down the data by focusing our experiments on the field of Computer Science. The field of each paper has been declared by openalex-topic-classification model which assigns a topic from a set of 4,516 different fields. Each topic is a node of a tree which at its higher level has been associated with a subfield. Same as the relation between topic and subfield, each subfield has a parent which indicates its field among 26 domains. Even after applying the Computer Science fileter, the dataset is quite large, therefore, for the sake of benchmarking, we further limit the scope by iterating and pruning the tree from top to find the most common subfields including papers related to Artificial Intelligence, Computer Vision and Pattern Recognition, Information Systems, and Human-Computer Interaction published between 2021 and 2024. For the sake of evaluation, we used the published papers in the last month i.e., March 2024 as our \textit{test set}.

As mentioned earlier, to maintain a manageable dataset size for academic researchers, we collected a total of 1,212,094 papers and 4,078,091 authors. Table \ref{tab:dataset_stat} presents the statistics of our collection. As shown in this table, the collection contains $|C_\tau| = 1,204,150$ papers with 4,065,252 total authors and 1,589,723 unique authors. We also have a test set of papers for which reviewers need to be found. 
To ensure the authors of the test set papers exist in $C_\tau$, we performed post-processing steps. We filtered out papers whose authors do not exist in $C_\tau$. As a result, we ended up with 7,944 papers, from an initial set of 9,771 papers, where at least one author exists in $C_\tau$. We establish that this is a fair evaluation approach, akin to evaluating an IR system with incomplete judgments—a well-studied and reasonable assumption \cite{buckley2004retrieval,sakai2008information,arabzadeh2022shallow,bigdeli2024evaluating,arabzadeh2023quantifying,arabzadeh2023adele,arabzadeh-clarke-2024-frechetm,alaofi2024generative}.
More details about our dataset and data processing is available in our GitHub\footnote{\url{https://github.com/sadjadeb/exHarmony}} repository.
 
\begin{table}[t]
\centering
\caption{Statistics of \texttt{exHarmony} collection and test set.}
\begin{tabular}{p{6em}p{6em}p{8em}p{8em}}
\hline \hline
 & Papers & Total Authors & Unique Authors \\ \hline
Collection & 1,204,150 & 4,065,252 & 1,589,723 \\
Test set & 7,944 & 23,919 & 21,703 \\
\hline \hline 
\end{tabular}
\label{tab:dataset_stat}
\end{table}

\subsection{Gold Standards}
To make \texttt{exHarmony} self-supervised for labels and ground truth, eliminating the need for human annotations, we define six different versions of gold standards based on varying assumptions about ideal reviewers.

\begin{enumerate}
    \item{\textbf{\texttt{exHarmony-Authors}}}: In this approach, we assume that the best reviewers for a paper are its own authors, as they have the most expertise in the topic. Hence, RAP is framed as retrieving the authors of a given paper based on its title and abstract. While this method is convenient, as noted earlier, not all early-career authors may be suitable for peer review. Therefore, we also introduce an `established authors' subset where an author is considered to be `established' if they have published $N$ papers within a specific timeframe. In \texttt{exHarmony}, established authors are those with at least 15 publications in the OpenAlex dataset.
    \item{\textbf{\texttt{exHarmony-Cite}}}: Since the previous assumption might be too restrictive, and relying solely on paper authors can lead to sparse gold standards, we propose a broader approach. In this case, we consider the authors of papers cited by the paper under review as potential reviewers. These cited authors may provide relevant expertise and can be considered valid matches for reviewing. Similar to the first approach, we include both `all authors' and `established authors' subsets.
\item{\textbf{\texttt{exHarmony-SimCite}}}: Not all cited papers may share the same topic or level of specificity as the paper being analyzed. To mitigate \textit{topic drift}, this subset focuses on authors of the most similar cited papers, i.e., those with high similarity to the main paper. To do this, we find top-10 similar cited papers using SPECTER \cite{cohan2020SPECTER} embeddings based on the papers title and abstracts. Similar to the previous two sets, this set is also curated for all authors and established authors.
\end{enumerate}

\section{Benchmarking}

For the sake of benchmarking, here we report performance of different models ranging from lexical-based matching to neural models on \texttt{exHarmony} benchmark. 

\subsection{Retrieval Baselines}
Since we redefine the task as a retrieval problem, we report a set of SOTA models on the task. 
For the sparse retrievers, we ran experiments with BM25 \cite{robertson1995okapi} 
retrievers both with expanded queries by RM3 as pseudo relevance feedback. These baselines were implemented using the Pyserini package \cite{lin2021pyserini}. The title and abstract of papers were considered as queries, and the previous works of authors were considered as individual documents indexed by Lucene.

Regarding neural embedding-based approaches, we leveraged two widely used static neural-based methods for the RAP task: one based on Word Mover's Distance (WMD) \cite{kusner2015word} and the other based on Doc2Vec \cite{le2014distributed}. 
Neural embedding-based approaches leverage dense vector representations of text to measure similarities.  In these two methods, the distance between the embedding representation of queries (the work for which we want to find reviewers) and the collection is measured. For WMD, we used pre-trained embeddings from Google News with 300 dimensions \cite{kusner2015word}. We trained the embeddings from scratch on our collection with 100 dimensions to capture the semantic meaning of the documents \cite{le2014distributed}.

\begin{table}[t]
\centering
\caption{Performance on \texttt{exHarmony-Authors}."R" denotes a ranking-specific adaptation of a model, fine-tuned for ranking and information retrieval.}
\label{tab:laraauthors}
\begin{tabular}{l|cc|cc|cc|cc|cc|cc}
\hline\hline
& \multicolumn{6}{c|}{\textbf{All Authors}} & \multicolumn{6}{c}{\textbf{Established Authors}} \\ 
  & \multicolumn{2}{c|}{\textbf{nDCG}} & \multicolumn{2}{c|}{\textbf{MAP}} & \multicolumn{2}{c|}{\textbf{Recall}} & \multicolumn{2}{c|}{\textbf{nDCG}} & \multicolumn{2}{c|}{\textbf{MAP}} & \multicolumn{2}{c}{\textbf{Recall}} \\ 
\textbf{Method} & @10 & @100 & @10 & @100 & @10 & @100 & @10 & @100 & @10 & @100 & @10 & @100 \\ \hline
BM25 & 0.096 & 0.116 & 0.079 & 0.084 & 0.106 & 0.168 & 0.098 & 0.118 & 0.081 & 0.086 & 0.116 & 0.186 \\ 
Doc2Vec & 0.010 & 0.021 & 0.008 & 0.010 & 0.013 & 0.051 & 0.010 & 0.022 & 0.008 & 0.010 & 0.015 & 0.057 \\ 
WMD & 0.061 & 0.069 & 0.051 & 0.053 & 0.065 & 0.091 & 0.062 & 0.071 & 0.052 & 0.054 & 0.070 & 0.102 \\ 
MiniLm & 0.083 & 0.099 & 0.069 & 0.072 & 0.090 & 0.143 & 0.083 & 0.100 & 0.070 & 0.073 & 0.097 & 0.157 \\ 
BERT & 0.098 & 0.116 & 0.081 & 0.085 & 0.105 & 0.167 & 0.098 & 0.117 & 0.081 & 0.085 & 0.114 & 0.183 \\ 
SciBERT & 0.100 & 0.119 & 0.082 & 0.086 & 0.110 & 0.172 & 0.103 & 0.122 & 0.085 & 0.089 & 0.122 & 0.192 \\ 
SPECTER & 0.103 & 0.125 & 0.084 & 0.088 & 0.113 & 0.186 & 0.105 & 0.127 & 0.086 & 0.091 & 0.124 & 0.208 \\ 
SciBERT-R & 0.101 & 0.118 & 0.083 & 0.087 & 0.109 & 0.167 & 0.102 & 0.119 & 0.084 & 0.088 & 0.118 & 0.184 \\ 
\hline\hline
\end{tabular}
\end{table}

\begin{table}[t]
\centering
\caption{Performance on \texttt{exHarmony-Cite}.}
\label{tab:laracite}

\begin{tabular}{l|cc|cc|cc|cc|cc|cc}
\hline\hline
& \multicolumn{6}{c|}{\textbf{All Authors}} & \multicolumn{6}{c}{\textbf{Established Authors}} \\ 
  & \multicolumn{2}{c|}{\textbf{nDCG}} & \multicolumn{2}{c|}{\textbf{MAP}} & \multicolumn{2}{c|}{\textbf{Recall}} & \multicolumn{2}{c|}{\textbf{nDCG}} & \multicolumn{2}{c|}{\textbf{MAP}} & \multicolumn{2}{c}{\textbf{Recall}} \\ 
 \textbf{Method} & @10 & @100 & @10 & @100 & @10 & @100 & @10 & @100 & @10 & @100 & @10 & @100 \\ \hline
BM25 & 0.187 & 0.110 & 0.018 & 0.031 & 0.023 & 0.078 & 0.163 & 0.099 & 0.015 & 0.027 & 0.021 & 0.074 \\ 
Doc2Vec & 0.018 & 0.018 & 0.001 & 0.002 & 0.002 & 0.017 & 0.016 & 0.017 & 0.001 & 0.002 & 0.002 & 0.016 \\ 
WMD & 0.100 & 0.053 & 0.009 & 0.013 & 0.012 & 0.035 & 0.089 & 0.049 & 0.008 & 0.011 & 0.011 & 0.034 \\ 
MiniLM-R & 0.137 & 0.080 & 0.014 & 0.021 & 0.018 & 0.056 & 0.120 & 0.072 & 0.012 & 0.019 & 0.017 & 0.054 \\ 
BERT-R & 0.183 & 0.106 & 0.018 & 0.030 & 0.024 & 0.075 & 0.159 & 0.096 & 0.016 & 0.026 & 0.022 & 0.072 \\ 
SciBERT & 0.138 & 0.075 & 0.013 & 0.020 & 0.017 & 0.053 & 0.121 & 0.069 & 0.011 & 0.017 & 0.016 & 0.052 \\ 
SPECTER & 0.178 & 0.107 & 0.017 & 0.029 & 0.022 & 0.077 & 0.157 & 0.098 & 0.015 & 0.025 & 0.021 & 0.075 \\ 
SciBERT-R & 0.176 & 0.099 & 0.018 & 0.028 & 0.022 & 0.069 & 0.154 & 0.089 & 0.015 & 0.024 & 0.021 & 0.066 \\ 
\hline\hline
\end{tabular}
\end{table}

\begin{table}[t]
\centering

\caption{Performance on \texttt{exHarmony-SimCite}.}
\label{tab:larasimcite}

\begin{tabular}{l|cc|cc|cc|cc|cc|cc}
\hline\hline
& \multicolumn{6}{c|}{\textbf{All Authors}} & \multicolumn{6}{c}{\textbf{Established Authors}} \\ 
\textbf{Method} & \multicolumn{2}{c|}{\textbf{nDCG}} & \multicolumn{2}{c|}{\textbf{MAP}} & \multicolumn{2}{c|}{\textbf{Recall}} & \multicolumn{2}{c|}{\textbf{nDCG}} & \multicolumn{2}{c|}{\textbf{MAP}} & \multicolumn{2}{c}{\textbf{Recall}} \\ 
& @10 & @100 & @10 & @100 & @10 & @100 & @10 & @100 & @10 & @100 & @10 & @100 \\ \hline
BM25 & 0.151 & 0.142 & 0.037 & 0.058 & 0.048 & 0.150 & 0.131 & 0.132 & 0.033 & 0.051 & 0.046 & 0.146 \\ 
Doc2Vec & 0.012 & 0.021 & 0.002 & 0.004 & 0.004 & 0.030 & 0.011 & 0.021 & 0.002 & 0.004 & 0.004 & 0.030 \\ 
WMD & 0.080 & 0.067 & 0.019 & 0.025 & 0.025 & 0.066 & 0.071 & 0.064 & 0.017 & 0.022 & 0.024 & 0.066 \\ 
MiniLM-R & 0.111 & 0.103 & 0.028 & 0.040 & 0.037 & 0.107 & 0.096 & 0.096 & 0.025 & 0.036 & 0.035 & 0.104 \\ 
BERT-R & 0.153 & 0.140 & 0.038 & 0.057 & 0.049 & 0.146 & 0.132 & 0.130 & 0.034 & 0.050 & 0.047 & 0.141 \\ 
SciBERT & 0.118 & 0.105 & 0.029 & 0.041 & 0.038 & 0.108 & 0.103 & 0.099 & 0.026 & 0.037 & 0.036 & 0.107 \\ 
SPECTER & 0.158 & 0.159 & 0.039 & 0.065 & 0.051 & 0.174 & 0.139 & 0.150 & 0.036 & 0.058 & 0.049 & 0.170 \\ 
SciBERT-R & 0.145 & 0.131 & 0.037 & 0.053 & 0.047 & 0.134 & 0.126 & 0.122 & 0.033 & 0.047 & 0.045 & 0.130 \\ 
\hline\hline
\end{tabular}
\end{table}

We further address the task using transformer-based contextualized neural embeddings. This group of retrievers, also known as dense retrievers, has shown excellent performance on downstream IR and natural language processing tasks \cite{arabzadeh2021bert}. We consider the following language models to solve the task. In general, these methods work by indexing the documents' dense vector representations using FAISS \cite{douze2024faiss}. During inference, the papers in the test set are first embedded, and then, using nearest neighbor search, authors with the most relevant and similar papers in the index are retrieved. Essentially, those whose embedding representations are most similar to the query, measured by cosine similarity, are retrieved \cite{reimers2019sentence}.

\begin{itemize}
    \item \textbf{Models trained for ranking:} We leveraged two pre-trained language models, MiniLm\footnote{\url{https://huggingface.co/sentence-transformers/msmarco-MiniLm-L6-cos-v5}} \cite{wang2020MiniLm} and BERT\footnote{\url{https://huggingface.co/sentence-transformers/msmarco-bert-base-dot-v5}} \cite{devlin2018bert}, both fine-tuned for the ranking task on the MS MARCO \cite{nguyen2016ms} dataset. MS MARCO is a widely used dataset for training and evaluating ranking tasks. Training on MS MARCO has proven to be effective and generalizable for different ad hoc retrieval tasks. These models were used to make the relevant pairs' representations closer together in the embedding space and irrelevant pairs further apart \cite{karpukhin2020dense,reimers2019sentence}.
    
    \item \textbf{Vanilla scientific-based language models:} To explore the impact of training on scientific literature, we leveraged two widely used language models for scientific purposes: SciBERT\footnote{\url{https://github.com/allenai/scibert}} \cite{beltagy2019scibert} and SPECTER\footnote{\url{https://github.com/allenai/SPECTER}} \cite{cohan2020SPECTER}. SciBERT is a BERT-based model pre-trained on a large corpus of scientific text from Semantic Scholar. It has shown high performance on various downstream tasks related to scientific content. SPECTER, on the other hand, is trained specifically to create document-level representations by learning to predict the co-citation context of papers. Both models have demonstrated strong results in scientific NLP tasks.
    
    \item \textbf{Scientific-based pre-training with ranking fine-tuning:} Lastly, we combined the previous categories by taking a language model pre-trained on scientific articles and fine-tuning it for the ranking task. Due to computational limitations, we only ran experiments with SciBERT. We aimed to fine-tune SciBERT for ranking. However, due to the lack of available datasets specifically for this, we fine-tuned it on the task of given a title, retrieving the corresponding abstract. The positive pairs for contrastive learning were titles and their related abstracts. We built negative pairs by randomly selecting from the top-1000 retrieved abstracts with BM25. Given the query, positive pair, and negative pairs, we trained a bi-encoder based dense retriever called fine-tuned SciBERT.
\end{itemize}
%
%

After retrieving the authors, we need to apply an aggregation function to remove redundant authors. To do this, we remove any repeated authors from the ranking by sorting the authors based on their relevance scores and then order by their frequency of number of papers in the retrieved list. We go down the retrieved list of reviewers and keep only the highest-ranked entry from each authors to remove redundant suggested reviewers.

\subsection{Evaluation}
In this section, we explain how \texttt{exHarmony} can be utilized for benchmarking, considering both the relevance of reviewers and the diversity of the reviewer set.

\subsubsection{Relevance}
To measure the effectiveness (based on \textit{relevance}) of RAP methods, we suggest using traditional IR evaluation metrics. These metrics reward a solution based on how well it ranks the reviewers of a given test paper. Specifically, we report Mean Average Precision (MAP), normalized Discounted Cumulative Gain (nDCG), and Recall, covering both precision-oriented and recall-oriented metrics. Additionally, we report these metrics at both shallow and deeper depths, i.e., for the top-10, top-20 and the top-100 retrieved reviewers.

\subsubsection{Diversity}
To measure the diversity of the set of retrieved reviewers, we propose the following metrics:

\begin{itemize}
    \item $\sigma_{\#cite}$:  It is essential to have a diverse set of reviewers ranging from junior to more senior researchers. This ensures \textit{inclusivity} and allows different perspectives. We measure the standard deviation of the number of citations among the set of top suggested reviewers at different depths. We refer to this as $\sigma_{\#cite}$.
    
    \item $\sigma_{\#paper}$: Since citations alone might be biased (e.g., one paper with many citations), we also explore the longevity of the reviewer's experience in the community. We consider the number of previous papers as an indicator of an reviewer's experience. The standard deviation of the number of published papers indicates diversity in experience: a higher standard deviation means greater diversity, while a lower standard deviation means the authors have a similar range of experience. Thus, a higher standard deviation is preferred.
    
    \item $\#Institutions$: To measure the diversity of the authors' backgrounds, we consider the number of unique institutions represented in the top-k retrieved reviewers. A higher number of unique institutions indicates a more diverse set of reviewers.
\end{itemize}

\subsection{Findings and Observations}
In this section, we report the performance of baselines in terms of \textit{relevance} as well as \textit{diversity}.

\subsubsection{Relevance Effectiveness.}
\begin{table*}[t]
\centering
\caption{Diversity Metrics for Different Retrieval Methods on \texttt{exHarmony}}
\label{tab:diversity_metrics}

\begin{tabular}{l|cc|cc|cc}
\hline\hline
& \multicolumn{2}{c|}{\textbf{$\sigma_{\#Cite}$}} & \multicolumn{2}{c|}{\textbf{$\sigma_{\#Paper}$}} & \multicolumn{2}{c}{\textbf{$\#Institutions$}} \\ 
\textbf{Method} & @10 & @100 & @10 & @100 & @10 & @100 \\ \hline
\textbf{BM25} & 17,262 & 28,990 & 1,021 & 1,722 & 7.31 & 66.83 \\ 
\textbf{Doc2Vec} & 15,460 & 29,385 & 872 & 1,678 & 7.67 & 71.63 \\ 
\textbf{WMD} & 15,260 & 26,984 & 903 & 1,602 & 7.27 & 67.78 \\ \hline
\textbf{MiniLm-R} & 16,463 & 28,936 & 989 & 1,721 & 7.32 & 68.05 \\ 
\textbf{BERT-R} & 16,086 & 28,087 & 960 & 1,678 & 7.32 & 67.86 \\ 
\textbf{SciBERT} & 15,884 & 26,242 & 939 & 1,542 & 7.31 & 66.65 \\ 
\textbf{SPECTER} & 17,080 & 27,073 & 999 & 1,598 & 7.33 & 67.20 \\ 
\textbf{SciBERT-R} & 15,669 & 26,909 & 922 & 1,581 & 7.29 & 67.26 \\ \hline\hline
\end{tabular}
\end{table*}

We first report nDCG, Recall, and MAP at cut-offs of 10, and 100 in Tables \ref{tab:laraauthors}--\ref{tab:larasimcite} for \texttt{exHarmony-Authors},\texttt{exHarmony-Cite} and \texttt{exHarmony-SimCite}, respectively. From these tables, we make the following observations: (1) Among the three groups of sparse, static-based, and contextualized-based neural embeddings, static-based neural embeddings perform the worst, showing poor performance. We argue that this is because the document lengths are relatively long, and methods like WMD and Doc2Vec are not able to capture the semantic and contextual information effectively. Additionally, due to specialized terminologies, these methods fail to create semantically meaningful representations. (2)  Lexical-based retriever, i.e., BM25 
show relatively good performance and are better than static-based neural methods in terms of both precision-oriented and recall-oriented metrics. The results and observations are consistent across all three subsets. (3) Across the contextualized-based neural embedding methods, those trained on scholarly literature, i.e., SciBERT and SPECTER, have shown better performance compared to general-purpose language models such as MiniLm and BERT that were trained on ranking tasks on MS MARCO. Among the scholarly-based language models, fine-tuning on the ranking task does not show significant improvement. This can be because due to the lack of available data, we trained the ranking task based on the given title and retrieved its abstract. This observation indicates that such an approach for fine-tuning might not be effective and may not lead to improvements.
\noindent (4) The performance range on \texttt{exHarmony-Authors} is lower than on \texttt{exHarmony-SimCite}, followed by \texttt{exHarmony-Cite}. We argue that this is because the number of gold standard reviewers in \texttt{exHarmony-Authors} is smaller than in \texttt{exHarmony-SimCite} and \texttt{exHarmony-Cite}. Having fewer gold standard reviewers makes the task more challenging.
\noindent (5) Additionally, we observe that sparse retrievers like BM25 perform better on \texttt{exHarmony-Cite} and \texttt{exHarmony-SimCite} compared to \texttt{exHarmony-Authors}. We hypothesize that this is because sparse retrievers rely on exact matching, which works well for finding similar papers with related topics in the scientific domain, where lexical matching is effective and semantic drift happens less frequently.

\subsubsection{Diversity.}
Now, we discuss the results of diversity metrics as reported in Table \ref{tab:diversity_metrics}. We note that higher $\sigma_{\#Cite}$, $\sigma_{\#Paper}$, and $\#Institutions$ indicate higher diversity in the set of proposed reviewers. From this table, we observe the following: (1) Static-based neural embedding methods (WMD and Doc2Vec) not only show poor performance in terms of relevance but also exhibit poor performance in terms of diversity. They show relatively lower standard deviations across all three diversity metrics. (2) Sparse retrievers like BM25, while demonstrating competitive performance in terms of relevance effectiveness, exhibit a relatively lower degree of diversity on all three metrics. (3) Among the contextualized-based approaches, SciBERT trained on a ranking task shows lower diversity, whereas SPECTER demonstrates the highest level of diversity. Generally, fine-tuning on ranking tasks seems to introduce biases into the model, which is understandable as more training can introduce more bias in the language model. (4) On average, across the top-100 suggested reviewers, there are approximately 65-70 unique institutions across all models. Doc2Vec, despite its poor performance, shows a high $\#Institutions$, which we hypothesize might actually be noise in the outcome.

\subsection{Discussions}
RAP remains a challenging task, as evidenced by the generally low performance metrics observed across various methods. This highlights the need for ongoing efforts to curate and maintain high-quality datasets and also possibly to redefine the task, as suggested in our approach. By utilizing authorship as a label, we can leverage the inherent expertise of authors, but even this method shows room for improvement. The performance of models, while indicative of potential, is still too low, underscoring the necessity for further refinement and optimization of these approaches.

Diversity is a critical factor that must be considered in RAP. Ensuring a diverse set of reviewers can provide multiple perspectives, enhance the quality of reviews, and promote inclusivity within the academic community. Our results demonstrate that while some methods, particularly contextualized neural embeddings, show promise in improving both relevance and diversity, there is still significant work to be done. It is essential to continue developing models and metrics that balance these aspects effectively. Addressing the challenges in RAP requires a multifaceted approach. Improving dataset quality, refining model performance, and prioritizing diversity are all essential steps toward creating a more effective and equitable peer review process. 
\section{Concluding Remarks}
In this paper, we have tackled the Reviewer Assignment Problem (RAP) by introducing the \texttt{exHarmony} benchmark. By leveraging the comprehensive data from OpenAlex, we proposed a novel approach that redefines RAP as a retrieval task, where an author is considered the best potential reviewer for their own paper. This approach addresses the challenge of obtaining gold standard data and provides a new framework for evaluating reviewer assignments based on relevance and diversity.

We provide the results of set of state of the art baselines on three subsets of \texttt{exHarmony-Authors}, \texttt{exHarmony-Cite} and \texttt{exHarmony-SimCite}. In addition, we have introduced new evaluation metrics that consider the diversity of the proposed reviewer set, ensuring inclusivity and a broader range of perspectives. Our work emphasizes the need for further exploration and more in-depth research to improve the efficiency and effectiveness of the reviewer assignment process.
For reproducibility and to facilitate future research, we have made all data and code publicly available. We hope that the \texttt{exHarmony} benchmark will serve as a valuable resource for the academic community and drive further advancements in this critical area of scholarly publishing.

\bibliographystyle{splncs04}
\bibliography{references}
\end{document}